\documentclass{article}
\usepackage{amssymb}
\usepackage{bm}
\usepackage[dvips]{graphicx}
\begin{document}

\title{Current-induced exchange switching magnetic junctions with cubic anisotropy of the free layer}

\author{S. G. Chigarev, E. M. Epshtein\thanks{E-mail: eme253@ms.ire.rssi.ru}, Yu. V. Gulyaev, P. E.
Zilberman\\ \\
V.A. Kotelnikov Institute of Radio Engineering and Electronics\\
    of the Russian Academy of Sciences, 141190 Fryazino, Russia}

\date{}

\maketitle

\abstract{The stability is analyzed of the equilibrium configurations of a
magnetic junction with a free layer that has cubic symmetry and two
anisotropy axes in the layer plane. Different variants of the switching between various
configurations are considered. A possibility is shown of the substantial
lowering of the threshold current density needed for the switching.
Numerical simulation is made of the switching dynamics for various
configurations.}

\section{Introduction}\label{section1}

Switching magnetic junctions by a spin-polarized current is one of the
main spintronic effects. Besides the academic interest, this phenomenon
may be used for high-density information processing, since the
characteristic scales are the exchange interaction and the spin diffusion
lengths of the order of tens of nanometers.

Nowadays interest has revived to magnetic junctions the layers of which
have cubic magnetic anisotropy; the well-studied iron may be an
example~\cite{Grabowski,Lehndorff,Wang}. Thin Fe(001) films have two equivalent anisotropy axes, [100] and
[010], in the layer plane. This allows switching the layer
magnetizations between different easy axes by means of magnetic field
and/or spin-polarized current, which may be used in memory cells with more
than two stable states.

In the present work, we consider a magnetic junction with cubic
anisotropy of the free layer placed between pinned and
nonmagnetic ones. Switching such systems by applied magnetic field has
been studied in Refs.~\cite{Lehndorff,Leonov}. Here we investigate the switching by
spin-polarized current at various relative orientations of the pinned and
free layer magnetization vectors. The thickness of the free layer is assumed to be small
compared to the spin diffusion length, so that the macrospin approximation
is valid~\cite{Gulyaev1}.

Two main mechanisms are known of the interaction between spin-polarized
current and magnetic lattice, namely, spin transfer torque (STT)~\cite{Slonczewski,Berger}
and spin injection leading to appearing regions of nonequilibrium spin
polarization near the interfaces~\cite{Heide,Gulyaev2} (the term ``field-like torque'' is
used sometimes in literature, because the mechanism action is equivalent
to influence of some effective magnetic field in some cases). The first of
the mechanisms mentioned (STT) is related with appearing a negative
damping that prevails over the positive Gilbert damping above the
threshold current density; this leads to instability of the original
magnetic configuration. The other (``injection'') mechanism is based on
increasing the \emph{sd} exchange interaction energy between the nonequilibrium
conduction electrons and the lattice under spin injection to the free
layer of the magnetic junction, so that the original state becomes
unstable, and a reorientation phase transition occurs. A unified theory of
switching magnetic junctions including actions of both mechanisms was
presented in Refs.~\cite{Gulyaev3,Elliott1}. The relative contribution of the mechanisms
indicated depends on the layer parameters and the applied magnetic field.
Earlier, the conditions were formulated~\cite{Gulyaev4,Epshtein1,Gulyaev1} under which the
injection mechanism plays the main role (see below for details). In this
work, special attention is paid to the latter.

The main equations describing the magnetization of the thin free layer in
the magnetic junction under spin-polarized current are presented in Sec.~\ref{section2}.
In Sec.~\ref{section3} we analyze the stability of the stationary configurations
depending on the current density through the magnetic junction, as well as
possible switching between these configurations. In Sec.~\ref{section4} the
current-voltage characteristic is found of the magnetic tunnel junction
under forward and backward currents depending on the original
configurations. In Sec.~\ref{section5} the junction switching dynamics is simulated
numerically.

\section{The main equations}\label{section2}
Let us consider a magnetic junction consisting of a pinned ferromagnetic
layer 1, a free ferromagnetic layer 2, and a nonmagnetic layer 3 which
closes the electric circuit. There is a thin spacer between the layers 1
and 2 that prevents the direct exchange interaction between the magnetic
lattices of the layers. The electric current flows perpendicular to the
layers (CPP mode). The free layer 2 has cubic symmetry with three mutually
orthogonal symmetry axes and, correspondingly, magnetic anisotropy
axes, one of which, [001], is perpendicular to the layer plane, while two
other, [100] and [010], lie in the plane. The anisotropy energy of that
layer (per area unit) is~\cite{Krinchik}
\begin{eqnarray}\label{1}
  &&U_a=\frac{1}{2}MH_aL\bigl\{\left(\hat\mathbf{M}\cdot\mathbf{n}_1\right)^2\left(\hat\mathbf{M}
  \cdot\mathbf{n}_2\right)^2+\left(\hat\mathbf{M}\cdot\mathbf{n}_2\right)^3\left(\hat\mathbf{M}
  \cdot\mathbf{n}_3\right)^2\nonumber\\
  &&+\left(\hat\mathbf{M}\cdot\mathbf{n}_3\right)^2\left(\hat\mathbf{M}
  \cdot\mathbf{n}_1\right)^2\bigl\},
\end{eqnarray}
where $L$ is the thickness of layer 2, $\mathbf M$ is the saturation magnetization of
that layer, $\hat\mathbf{M}=\mathbf M/|M|$ is the unit vector along the magnetization,
$\mathbf n_1$, $\mathbf n_2$, $\mathbf n_3$ are the unit vectors along [100], [010] and [001] axes,
respectively, $H_a$ is the effective anisotropy field.

The free layer thickness $L$ is assumed to be small compared to the spin
diffusion length $l$ and the inhomogeneity scale of the magnetic lattice in
that layer (such a scale, the measure of the ``spatial inertia'' of the
lattice, is the domain wall thickness $\delta$). Under such conditions, layer 2
manifests itself as united whole (``macrospin'') in respect of its magnetic
behavior. This leads to the modification of the Landau--Lifshitz--Gilbert
equation for the layer magnetization with the disappearance of the
spatial derivative term and the introduction of a new term describing the
current effects. With the cubic anisotropy taken into account, the equation
takes the form (cf.~\cite{Gulyaev1})
\begin{eqnarray}\label{2}
  &&\frac{d\hat\mathbf M}{dt}-\kappa\left(\hat\mathbf M\times\frac{d\hat\mathbf M}{dt}\right)
  +\gamma\left(\hat\mathbf M\times\mathbf H\right)+\gamma\left(\hat\mathbf M\times\mathbf H_d\right)\nonumber\\
&&-\gamma H_a\Bigl\{\left(\hat\mathbf M\cdot\mathbf n_1\right)\left(\hat\mathbf M\cdot\mathbf n_2\right)
\Bigl\{\left(\hat\mathbf M\cdot\mathbf n_2\right)\left(\hat\mathbf M\times\mathbf n_1\right)\nonumber\\
&&+\left(\hat\mathbf M\cdot\mathbf n_1\right)\left(\hat\mathbf M\times\mathbf n_2\right)\Bigr\}
+\left(\hat\mathbf M\cdot\mathbf n_2\right)\left(\hat\mathbf M\cdot\mathbf n_3\right)
\Bigl\{\left(\hat\mathbf M\cdot\mathbf n_3\right)\left(\hat\mathbf M\times\mathbf n_2\right)\nonumber\\
&&+\left(\hat\mathbf M\cdot\mathbf n_2\right)\left(\hat\mathbf M\times\mathbf n_3\right)\Bigr\}
+\left(\hat\mathbf M\cdot\mathbf n_3\right)\left(\hat\mathbf M\cdot\mathbf n_1\right)
\Bigl\{\left(\hat\mathbf M\cdot\mathbf n_1\right)\left(\hat\mathbf M\times\mathbf n_3\right)\nonumber\\
&&+\left(\hat\mathbf M\cdot\mathbf n_3\right)\left(\hat\mathbf M\times\mathbf
n_1\right)\Bigr\}\Bigr\}
\nonumber\\
&&=-\frac{a}{L}\left\{p(\hat\mathbf M)\left(\hat\mathbf M\times\hat\mathbf M_1\right)
+\kappa(\hat\mathbf M)\left(\hat\mathbf M\times\left(\hat\mathbf M\times\hat\mathbf M_1\right)\right)\right\}.
\end{eqnarray}
Here $\hat\mathbf M_1$ is the unit vector along the magnetization of layer 1, $\mathbf H$ is the
applied magnetic field, $\mathbf H_d$  is the demagnetization field, $\kappa$ is the Gilbert
damping factor, $\gamma$ is the gyromagnetic ratio, $a=\gamma H_a\delta^2$ is the lattice
magnetization diffusion constant,
\begin{eqnarray}\label{3}
  &&p(\hat\mathbf M)=\frac{\mu_B\gamma\alpha\tau Q_1}{ea}j\lambda\frac{Z_1}{Z_2}\Bigl[\frac{Z_1}{Z_3}
  +\frac{Z_1}{Z_2}\lambda-\left(\hat\mathbf M_1\cdot\hat\mathbf M\right)^2\nonumber\\
  &&+\frac{2b}{\lambda}\left(\lambda
  +\frac{Z_2}{Z_3}\right) \left(\hat\mathbf M_1\cdot\hat\mathbf M\right)\Bigr]\nonumber\\
&&\times\left[\frac{Z_1}{Z_3}+\frac{Z_1}{Z_2}\lambda+\left(\hat\mathbf M_1\cdot\hat\mathbf M\right)^2\right]^{-2},
\end{eqnarray}

\begin{equation}\label{4}
  k(\hat\mathbf
  M)=\displaystyle\frac{\mu_BQ_1}{eaM}j\left(\frac{Z_1}{Z_3}+\frac{Z_1}{Z_2}\lambda\right)
\left[\frac{Z_1}{Z_3}+\frac{Z_1}{Z_2}\lambda+\left(\hat\mathbf M_1\cdot\hat\mathbf M\right)^2\right]^{-2},
\end{equation}
where $e$ is the electron charge, $\mu_B$ is the Bohr magneton, $\alpha$ is the
\emph{sd} exchange interaction constant, $\tau$ is the spin relaxation time, $\lambda=L/l\ll1$, $Q$ is the
conduction spin polarization,
\begin{equation}\label{5}
  Z_i=\displaystyle\frac{\rho_il_i}{1-Q_i^2}\qquad(i=1,\,2,\,3)
\end{equation}
is the layer spin resistance~\cite{Epshtein1}, $\rho$ is the electric resistivity; the
quantities without index refer to the free layer 2. The $b=(\alpha_1M_1\tau_1)/(\alpha M\tau)$ parameter
describes influence of the pinned layer 1.

The parameters $p$ and $k$ in the right-hand side of Eq.~(\ref{2}) describe
injection and STT mechanisms of the spin-polarized current effect on the
magnetic lattice under current flowing in the ``forward'' direction,
corresponding to the electron drift in the $1\to2\to3$ direction. A
substitution $\left(\hat\mathbf M_1\cdot\hat\mathbf M\right)\to
\left(\hat\mathbf M_1\cdot\hat\mathbf M\right)^{-1}$ , $j\to-|j|$ corresponds
to the backward direction $(3\to2\to1)$.

As analysis shows~\cite{Gulyaev4}, the intensity of the spin injection to the free
layer is determined with the relationship between the spin resistances of
the layers. At $\displaystyle\frac{Z_2}{Z_1},\,\displaystyle\frac{Z_2}{Z_3}\ll\lambda\ll1$ relationship
between the spin resistances of the pinned
layer 1, free layer 2 and nonmagnetic layer 3, high spin injection level
is reached due to effective injection from layer 1 to layer 2 and ``locking
up'' the injection from layer 2 to layer 3~\cite{Epshtein1}. This leads to lowering the
switching threshold due to the injection mechanism. Another possibility of
the threshold lowering consists in applying external magnetic field~\cite{Gulyaev1}.
The STT contribution becomes unimportant under low threshold of the
injection switching magnetic junction, so that we will consider only the
injection mechanism below.

Under assumptions mentioned, the Landau--Lifshitz equation describing the
lattice dynamics of the free layer in the absence of damping has an
integral of motion even in presence of a current through the junction~\cite{Elliott2,Epshtein2}.
Such an integral is the magnetic energy, including the Zeeman energy
in applied magnetic field, the anisotropy energy, the demagnetization
energy, and the \emph{sd} exchange interaction energy between conduction
electrons and magnetic lattice.

Let us consider a configuration with $x$ axis along the current, $yz$ plane
parallel to the layer planes $\mathbf H=\{0,\,H\sin\psi,\,H\cos\psi\}$,
$\mathbf n_1=\{0,\,0,\,1\}$, $\mathbf n_2=\{0,\,1,\,0\}$, $\mathbf n_3=\{0,\,0,\,1\}$, $\mathbf H_d=-4\pi M\{\hat
M_x,\,0,\,0\}$, $\hat\mathbf M_1=\{0,\,0,\,1\}$.

In spherical coordinates with the polar axis along [100] axis,
$\hat\mathbf M=\{\sin\theta\cos\phi,\,\sin\theta\sin\phi,\,\cos\theta\}$, the
dimensionless (in $MH_aL$ units) magnetic energy under forward current takes the
form
\begin{eqnarray}\label{6}
  &&\frac{U(\theta,\,\phi)}{MH_aL}=-\frac{H}{H_a}\cos(\theta-\psi)+\frac{1}{2}\sin^2\theta\cos^2\theta
  +\frac{1}{2}\sin^4\theta\sin^2\phi\cos^2\phi\nonumber\\
&&+\frac{2\pi M}{H_a}\sin^2\theta\cos^2\phi
-\frac{j}{j_0}\frac{\cos\theta+b(Z_2/Z_1\lambda)\cos^2\theta}{1+(Z_2/Z_3\lambda)
+ (Z_2/Z_1\lambda)\cos^2\theta},
\end{eqnarray}
where $j_0=\displaystyle\frac{eH_aL}{\mu_B\alpha\tau Q_1}$.

The corresponding formula for backward current is obtained with
substitution $\cos\theta\to(\cos\theta)^{-1}$, $j\to-|j|$ in the last term of Eq.~(\ref{6})
describing the current effect:
\begin{eqnarray}\label{7}
  &&\frac{U(\theta,\,\phi)}{MH_aL}=-\frac{H}{H_a}\cos(\theta-\psi)+\frac{1}{2}\sin^2\theta\cos^2\theta
  +\frac{1}{2}\sin^4\theta\sin^2\phi\cos^2\phi\nonumber\\
&&+\frac{2\pi M}{H_a}\sin^2\theta\cos^2\phi
+\frac{j}{j_0}\frac{\cos\theta+b(Z_2/Z_1\lambda)}{
(Z_2/Z_1\lambda)+[1+(Z_2/Z_3\lambda)]\cos^2\theta}.
\end{eqnarray}

\section{Stationary states and variants of switching}\label{section3}
The stationary states of the system in study correspond to the extrema of
the $U(\theta,\,\phi)$ function; the minima of the function correspond to the stable
equilibrium states. Because of the positive definiteness of the term with
the azimuthal angle $\phi$, it is sufficient to consider only the energy
dependence on the polar angle $\theta$ at fixed value $\phi=90^\circ$ during the minima finding
(this corresponds to in-plane position of the magnetization vector).

There are three minima in absence of magnetic field ($H=0$) and current
($j=0$): $\theta=0^\circ$, $\theta=90^\circ$, and $\theta=180^\circ$, corresponding to the
parallel, perpendicular and antiparallel
relative orientations of the pinned and free layers.

\begin{figure}
\includegraphics{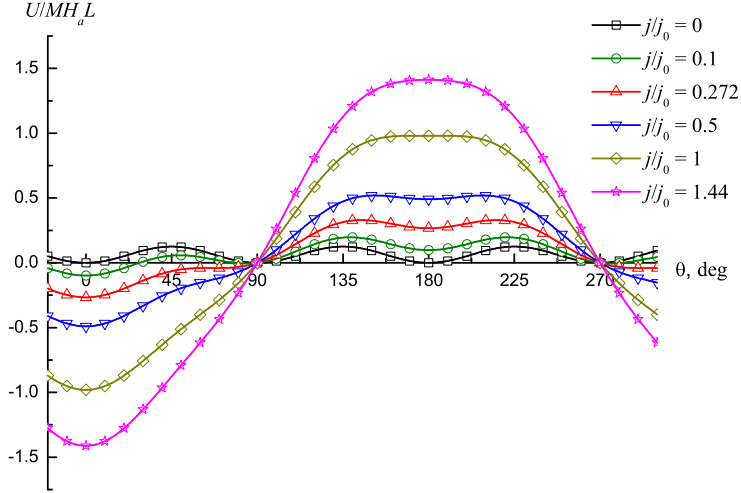}
\caption{The magnetic junction energy as a function of the angle between
the magnetization vectors of the pinned and free layers with various
values of the (dimensionless) forward current density.}\label{fig1}
\end{figure}

\begin{figure}
\includegraphics{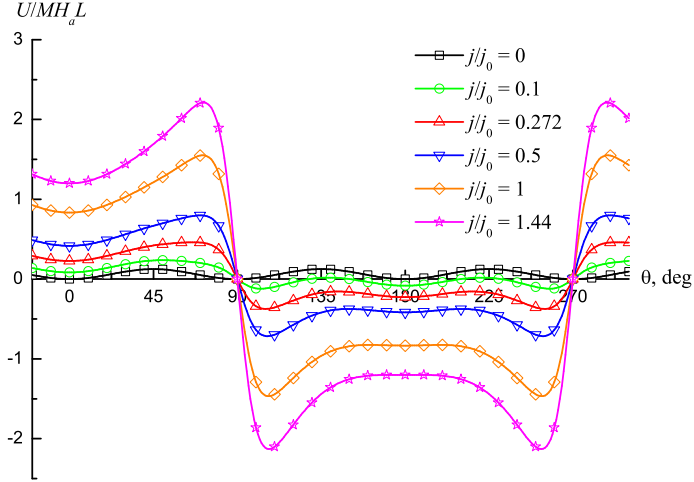}
\caption{The magnetic junction energy as a function of the angle between
the magnetization vectors of the pinned and free layers with various
values of the (dimensionless) backward current density.}\label{fig2}
\end{figure}

Let us present the stability analysis results for these stationary states
in presence of the current under intense injection
conditions $\displaystyle\frac{Z_2}{Z_1},\,\displaystyle\frac{Z_2}{Z_3}\ll\lambda\ll1$ (as
mentioned above, the latter condition allows neglecting the STT
contribution). The results are illustrated with plots of the $U$ energy as a
function of the $\theta$ angle (Figs.~\ref{fig1} and~\ref{fig2}) at various values of the current
density in forward and backward directions.

1) The parallel configuration ($\theta=0^0$) in absence of magnetic field
($H=0$) is stable under forward and backward currents.

2) Under the forward current increasing from zero, the energy minimum
corresponding to the original perpendicular configuration ($\theta=90^\circ$) shifts in
the direction of the parallel configuration (i.e., $\theta$ angle decreases).
This takes place up to the current density value $j=j_0\sqrt{2/27}\approx0.272j_0$, when the deviation
from the original position reaches $\arcsin\left(1/\sqrt6\right)\approx24^\circ$
(i.e., $\theta\approx66^\circ$ ). At this value, the energy
minimum disappears (it changes to an inflection point), and the system
switches abruptly to a parallel configuration and remains in that
configuration under further variations of the current density, so that the
$90^\circ\to0^\circ$ switching is irreversible.

Under backward current increasing from zero, the minimum corresponding to
the original perpendicular configuration ($\theta=90^\circ$) shifts in the direction of
the antiparallel configuration, and the corresponding $\theta$ angle increases up
to some value $\theta_0(j)$ that tends to
\begin{equation}\label{8}
  \theta_0(\infty)=\arccos\left(-\sqrt{\frac{Z_2/Z_1\lambda}{1+( Z_2/Z_3\lambda)}}\right)
\end{equation}
at the high current limit. Under returning to zero current, the
perpendicular orientation restores, so that the switching by the backward
current is of ``temporary'' character.

3) The antiparallel configuration ($\theta=180^\circ$) in a magnetic field parallel to the
magnetization of the pinned layer ($\psi=0^\circ$) becomes unstable and switched to
parallel one under high enough forward current. The threshold current density is
\begin{equation}\label{9}
  j_{\rm{th}}=j_0\frac{\left[1+(Z_2/Z_3\lambda)+(Z_2/Z_1\lambda)\right]^2}
  {1+(Z_2/Z_3\lambda)-(Z_2/Z_1\lambda)}\left(1-\frac{H}{H_a}\right)\approx j_0
  \left(1-\frac{H}{H_a}\right).
\end{equation}

We see from Eq.~(\ref{9}) the mentioned possibility of the switching threshold
lowering by means of an applied magnetic field close to (but lower than)
the anisotropy field. Such an assistance of the magnetic field does not
break the local character of the switching, because the magnetic field
lower than the anisotropy field cannot do switching alone (without a
current).

The parallel configuration appeared after switching is stable against
further variations of the current, so that the switching by the forward
current is irreversible.

Under the backward current, the antiparallel configuration becomes
unstable at the same (in magnitude) current density $|j|=j_{\rm{th}}$, however, in this
case switching takes place to a nonequilibrium stationary state $\theta=\theta_0(j_{\rm{th}})$
or to the symmetrical state  $\theta=360^\circ-\theta_0(j_{\rm{th}})$.
With returning to zero current, the system does
not return to antiparallel configuration, but comes to one of two
perpendicular configurations $\theta=90^\circ$ or $\theta=270^\circ$.

Thus the following variants are possible of the switching between
stationary states: 1) switching an antiparallel configuration to a
parallel one by turning up the forward current of $j>j_{\rm{th}}$ density and subsequent
turning off; 2) switching an antiparallel configuration to a perpendicular
one by turning up the backward current of the same density and subsequent
turning off; 3) switching a perpendicular configuration to a parallel one
by turning up and subsequent turning off the forward current of
substantially lower density $j>0.272j_0$.

\section{Resistance of the magnetic tunnel junction}\label{section4}

\begin{figure}
\includegraphics{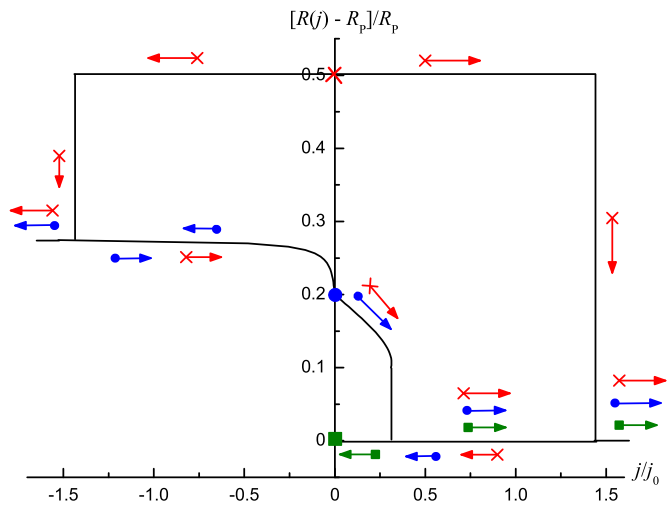}
\caption{The magnetic tunnel junction resistance as a function of the
(dimensionless) current density. The dots on the ordinate show the
stationary configurations without current: parallel (square),
perpendicular (circle), and antiparallel (skew cross). The arrows with
corresponding tails show the resistance changes under the current change
for different initial configurations.}\label{fig3}
\end{figure}

In experiments, the current-driven switching magnetic junction manifests
itself, in the first place, as a change of the junction resistance. The
resistance depends substantially on the relative orientation of the layers
forming the junction; this is the cause of the well-known tunnel
magnetoresistance effect.

The conductance of a magnetic tunnel junction with $\theta$ angle between the
magnetization vectors of the layers takes the form~\cite{Utsumi}
\begin{equation}\label{10}
  G(\theta)=G_P\cos^2\frac{\theta}{2}+ G_{AP}\sin^2\frac{\theta}{2},
\end{equation}
where $G_P,\,G_{AP}$ are the junction conductances at parallel ($\theta=0^\circ$) and
antiparallel ($\theta=180^\circ$) relative orientation of the layers, respectively.

It is convenient to describe the change of the junction resistance with the following ratio:
\begin{equation}\label{11}
  \frac{R(\theta)-R_P}{R_P}=\frac{\rho(1-\cos\theta)}{2+\rho(1+\cos\theta)},
\end{equation}
where $R(\theta)=1/G(\theta)$, $R_P=1/G_P$; $\rho=[R(180^\circ)-R_P]/R_P$
is the tunnel magnetoresistance defined by usual way~\cite{Gulliere}.

To find the resistance dependence on the current direction and density $R(j)$,
it is necessary to substitute $\theta(j)$ dependence to Eq.~(\ref{11}). With the foregoing
analysis taking into account, the results are obtained shown in Fig.~\ref{fig3}. A
possibility is seen of the switching between different stationary states
corresponding to different electric resistances.

\section{Simulation of the magnetic junction switching dynamics}\label{section5}
Together with the investigation of the stationary states and the switching
between them, the switching dynamics is of great interest, because it
determines the speed of response of the devices based on the magnetic
junctions.

\begin{figure}
\includegraphics{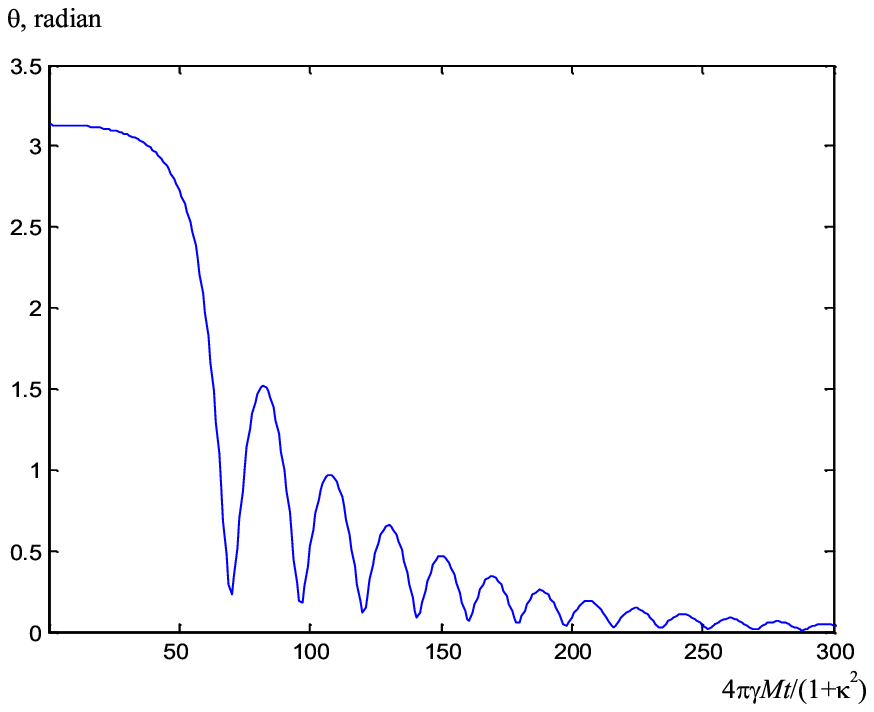}
\caption{Switching dynamics of the antiparallel configuration to parallel
one by the forward current.}\label{fig4}
\end{figure}

\begin{figure}
\includegraphics{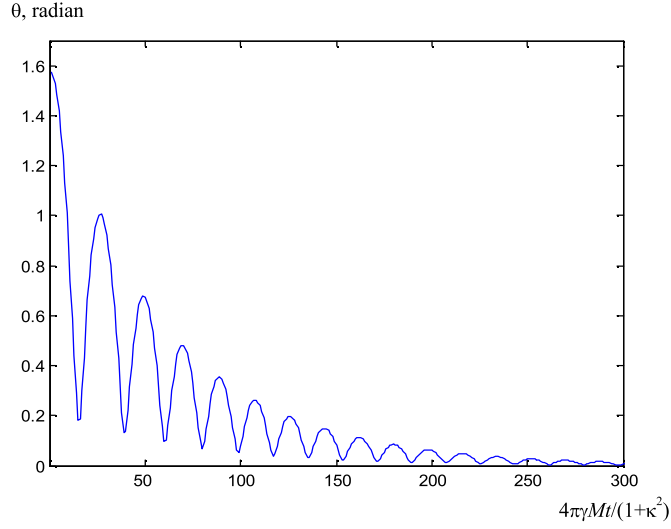}
\caption{Switching dynamics of the perpendicular configuration to parallel
one by the forward current.}\label{fig5}
\end{figure}

\begin{figure}
\includegraphics{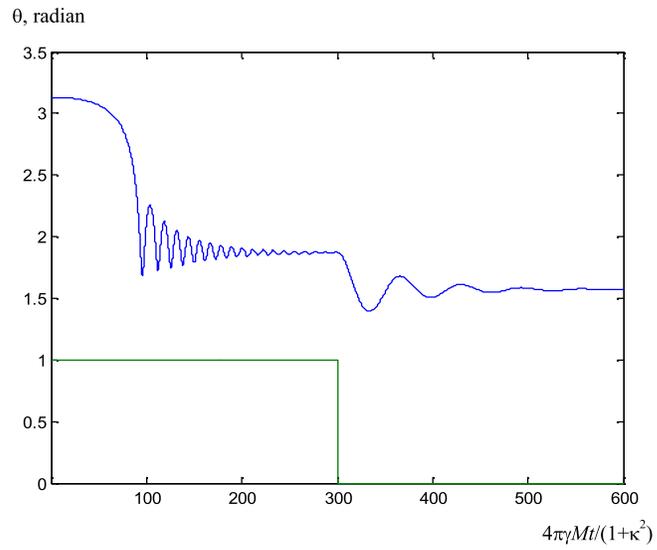}
\caption{Switching dynamics of the antiparallel configuration to
perpendicular one by the backward current. The step shows the current
turning-off time.}\label{fig6}
\end{figure}

\begin{figure}
\includegraphics{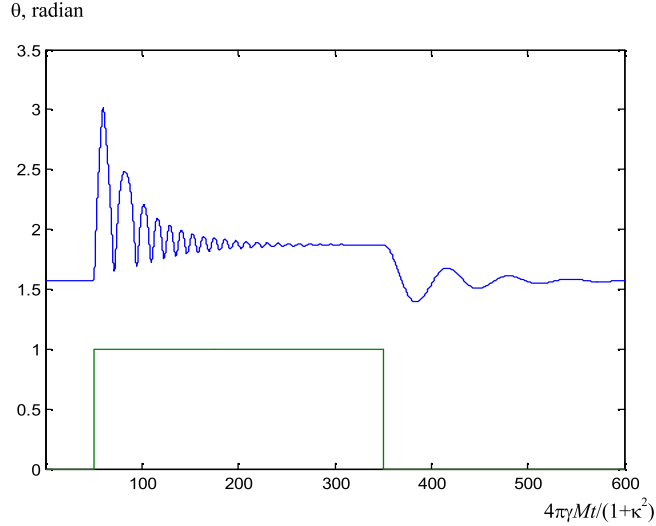}
\caption{The perpendicular configuration evolution under turning on and
turning off a rectangular pulse of the backward current.}\label{fig7}
\end{figure}

The time-dependent vector equation~(\ref{2}) describing the dynamics with using
polar coordinates $(\theta,\,\phi)$ takes the form of a set of equations
\begin{equation}\label{12}
  \frac{d\theta}{dT}=\frac{\sin\theta}{1+\kappa^2}\{-\kappa A(\theta,\,\phi)+B(\theta,\,\phi)\},
\end{equation}

\begin{equation}\label{13}
  \frac{d\phi}{dT}=\frac{1}{1+\kappa^2}\{ A(\theta,\,\phi)+ \kappa B(\theta,\,\phi)\},
\end{equation}
where
\begin{equation}\label{14}
  A(\theta,\,\phi)=h\cos\psi+h_a\cos\theta\cos2\theta+\cos\theta\cos^2\phi+P(\theta),
\end{equation}

\begin{equation}\label{15}
  B(\theta,\,\phi)=\cos\phi\sin\phi-K(\theta),
\end{equation}

\begin{displaymath}
  h=\frac{H}{4\pi M},\quad h_a=\frac{H_a}{4\pi M},\quad T=4\pi\gamma Mt.
\end{displaymath}

The $P(\theta)$ and $K(\theta)$ functions describing the current effect take the form
\begin{eqnarray}\label{16}
  &&P(\theta)=\frac{j}{j_0}\lambda h_a\frac{Z_1}{Z_2}\left[\frac{Z_1}{Z_3}+\frac{Z_1}{Z_2}\lambda-\cos^2\theta+\frac{2b}{\lambda}\left(\lambda
+\frac{Z_2}{Z_3}\right)\cos\theta\right]\nonumber\\
&&\times\left(\frac{Z_1}{Z_3}+\frac{Z_1}{Z_2}\lambda+\cos^2\theta\right)^{-2},
\end{eqnarray}

\begin{equation}\label{17}
  K(\theta)=\frac{j}{j_0}\frac{h_a}{\alpha\tau\gamma M}\left(\frac{Z_1}{Z_3}
  +\frac{Z_1}{Z_2}\lambda\right) \left(\frac{Z_1}{Z_3}
  +\frac{Z_1}{Z_2}\lambda+\cos^2\theta \right)^{-1}
\end{equation}
for the forward current, and
\begin{eqnarray}\label{18}
  &&P(\theta)=\frac{j}{j_0}\lambda h_a\frac{Z_1}{Z_2}\left[1-\left(\frac{Z_1}{Z_3}
  +\frac{Z_1}{Z_2}\lambda\right)\cos^2\theta-\frac{2b}{\lambda}\left(\lambda
+\frac{Z_2}{Z_3}\right)\cos\theta\right]\nonumber\\
&&\times\left[1+\left(\frac{Z_1}{Z_3}+\frac{Z_1}{Z_2}\lambda\right)\cos^2\theta\right]^{-2},
\end{eqnarray}

\begin{equation}\label{19}
  K(\theta)=-\frac{j}{j_0}\frac{h_a}{\alpha\tau\gamma M}\left(\frac{Z_1}{Z_3}
  +\frac{Z_1}{Z_2}\lambda\right)\cos^2\theta \left[1+\left(\frac{Z_1}{Z_3}
  +\frac{Z_1}{Z_2}\lambda\right)\cos^2\theta\right]^{-1}
\end{equation}
for the backward current.

Since $h_a\ll1$ in fact, the precession of the free layer magnetization
vector takes place in the layer plane on the whole, i.e., the vector end
describes an ellipse elongated strongly in the plane. So the small terms
proportional to $h_a\cos^2\phi$ were omitted in derivation of
Eqs.~(\ref{12})--(\ref{15}).

In Ref.~\cite{Epshtein2} an analytical solution was found of Eqs.~(\ref{12}),~(\ref{13})
at $K=0,\,h\ll1,\,h_a\ll1$ in
the limiting cases of the zero damping ($\kappa=0$) and strong damping ($\kappa\gg1$). In the
general case, it is reasonable to use numerical simulation. We used the
Simulink software of the MATLAB system that is intended for simulating
dynamical systems~\cite{Karris}.

The following parameter values were
given: $\kappa=0.03,\,h_a=0.01,\,h=0,\,\lambda=0.1,\,Z_2/Z_1
=Z_2/Z_3=0.01,\,\alpha\tau\gamma M=60$ (cf.~\cite{Gulyaev1}). The
thermal noises initiating deviation of the free layer magnetization from
the original unstable equilibrium state were imitated with giving a small
initial deviation from such a state by an angle of 0.01 radians in the layer
plane where the demagnetization field does not prevent fluctuation-induced
deviations, so that minimal fluctuation energy is needed.
The time dependences of the magnetization deviation from [100] axis
$\theta(T)$ under forward current of $j=2j_0$ density for antiparallel and perpendicular
original configuration, respectively, are shown in Figs.~\ref{fig4} and~\ref{fig5}, the
similar ones for the backward current are shown in Figs.~\ref{fig6} and~\ref{fig7}. The
dimensionless time is laid off as abscissa
with $t_0=(1+\kappa^2)/(4\pi\gamma M)$ as the time unit; at $M=900$ G
one nanosecond corresponds to 200 scale divisions of the abscissa. The
steps in Figs.~\ref{fig6} and~\ref{fig7} show the current turning on and turning off times.

The numerical solution (simulation) results consist completely with
foregoing analysis based on the angular dependence of the magnetic energy.
It is seen that direct switching occurs of the antiparallel and
perpendicular configurations to parallel one under the forward current.
Under the backward current, the switching occurs to an intermediate
nonequilibrium stationary state, from which a transition (or return, when
the initial configuration is perpendicular) occurs to the perpendicular
state. As was to be expected, the switching is accompanied with damped
oscillation due to precession of the magnetization vector.

With given parameter values, the characteristic switching times are of the
order of nanoseconds, while the oscillation period is of the order of
fractions of nanosecond. With increasing the magnetization and the damping
constant, the speed of response rises (the latter up to some limits,
because the switching process becomes aperiodic and slows with increasing
damping at too strong damping ($\kappa\gg1$)).

\section{Conclusion}\label{section6}
The analysis shows a possibility of increasing the number of the
switchable states by using magnetic junctions with cubic-anisotropy layers.
The fact is of interest that the switching of the perpendicular
configuration to the parallel one requires current density lower by
several times, than the switching of the antiparallel configuration. The
fruitfulness should be noted of the combination of the energy approach to
determining stationary states with numerical simulation of the switching
processes.

\section*{Acknowledgments}
The authors are grateful to Yu. G. Kusraev, N. A. Maksimov and G. M. Mikhailov for useful discussions.

The work was supported by the Russian Foundation for Basic Research, Grant
No.~08-07-00290.


\begin{thebibliography}{12}
\bibitem{Grabowski}
J. Grabowski, M. Przybylski, M. Nyvlt, and J. Kirschner, J. Appl. Phys. \textbf{104}, 113905 (2008).
\bibitem{Lehndorff}
R. Lehndorff, M. Buchmeier, D. E. B\"urger D.E., et al., Phys. Rev. B \textbf{76}, 214420 (2007).
\bibitem{Wang}
S. G. Wang, R. C. C. Ward, G. X. Du, et al., Phys. Rev. B \textbf{78}, 180411 (2008).
\bibitem{Leonov}
A. A. Leonov, U. K. R\"{o}\ss ler, and A. N. Bogdanov, J. Appl. Phys. \textbf{104}, 084304 (2008).
\bibitem{Gulyaev1}
Yu. V. Gulyaev, P. E. Zilberman, A. I. Panas, and E. M. Epshtein, Zh.
Eksp. Teor. Fiz. \textbf{134}, 1200 (2008) [JETP \textbf{107}, 1027 (2008)].
\bibitem{Slonczewski}
J. C. Slonczewski, J. Magn. Magn. Mater. \textbf{159}, L1 (1996).
\bibitem{Berger}
L. Berger, Phys. Rev. B \textbf{54}, 9353 (1996).
\bibitem{Heide}
C. Heide, P. E. Zilberman, and R. J. Elliott, Phys. Rev. B \textbf{63}, 064424 (2001).
\bibitem{Gulyaev2}
Yu. V. Gulyaev, P. E. Zilberman, E. M. Epshtein, and R. J. Elliott,
Pis'ma Zh. Eksp. Teor. Fiz. \textbf{76}, 189 (2002) [JETP Letters \textbf{76}, 155 (2002)].
\bibitem{Gulyaev3}
Yu. V. Gulyaev, P. E. Zilberman, E. M. Epshtein, and R. J. Elliott, Zh.
Eksp. Teor. Fiz. \textbf{127}, 1138 (2005) [JETP \textbf{100}, 1005 (2005)].
\bibitem{Elliott1}
R. J. Elliott, E. M. Epshtein, Yu. V. Gulyaev, and P.E. Zilberman, J. Magn. Magn. Mater. \textbf{300}, 122 (2006).
\bibitem{Gulyaev4}
Yu. V. Gulyaev, P. E. Zilberman, A. I. Krikunov, and E. M. Epshtein, Zh.
Tekhn. Fiz. \textbf{77}, No. 9, 67 (2007) [Techn. Phys. \textbf{52}, 1169 (2007)].
\bibitem{Epshtein1}
E. M. Epshtein, Yu. V. Gulyaev, and P. E. Zilberman, J. Magn. Magn. Mater. \textbf{312}, 200 (2007).
\bibitem{Krinchik}
G. S. Krinchik, Physics of Magnetic Phenomena (Moscow State Univ. Publ., 1976) [in Russian].
\bibitem{Elliott2}
R. J. Elliott, E. M. Epshtein, Yu. V. Gulyaev, and P. E. Zilberman, J. Magn. Magn. Mater. \textbf{271}, 88
(2004).
\bibitem{Epshtein2}
E. M. Epshtein, Radiotekh. Elektron. (Moscow) \textbf{54}, 339 (2009) [J. Commun. Technol. Electron.
\textbf{54}, 323 (2009)].
\bibitem{Utsumi}
Y. Utsumi, Y. Shimizu, and H. Miyazaki, J. Phys. Soc. Japan \textbf{68}, 3444 (1999).
\bibitem{Gulliere}
M. Gulliere, Phys. Lett. A \textbf{54}, 225 (1975).
\bibitem{Karris}
S. T. Karris, Introduction to Simulink with Engineering Applications (Orchard
Publications, 2006).
\end{thebibliography}
\end{document}